\documentclass{article}

\usepackage{PRIMEarxiv}
\usepackage{xcolor}
\usepackage[utf8]{inputenc} 
\usepackage[T1]{fontenc}    
\usepackage{hyperref}       
\usepackage{url}            
\usepackage{booktabs}       
\usepackage{amsfonts}       
\usepackage{nicefrac}       
\usepackage{microtype}      
\usepackage{float}
\usepackage{lipsum}
\usepackage{fancyhdr}       
\usepackage{graphicx}       
\usepackage{amssymb,mathtools}  
\usepackage{array}
\usepackage{bm} 
\usepackage{caption}
\usepackage{subcaption}

\newcommand{\vect}[1]{\bm{#1}}

\title{Learning-Based Data-Assisted Port-Hamiltonian Control for Free-Floating Space Manipulators
}

\author{
  Mostafa Eslami, Maryam Babazadeh \\
  Department of Mechanical and Process
Engineering\\
RPTU Kaiserslautern-Landau\\
67663 Kaiserslautern, Germany,\\
\tt\small 
  \texttt{\{m.eslami, m.babazadeh\}rptu@de} 
}

\begin{document}
\maketitle

\begin{abstract}
A generic data-assisted control architecture within the port-Hamiltonian framework is proposed, introducing a physically meaningful observable that links conservative dynamics to all actuation, dissipation, and disturbance channels. A robust, model-based controller combined with a high-gain decentralized integrator establishes large robustness margins and strict time-scale separation, ensuring that subsequent learning cannot destabilize the primary dynamics. Learning, selected for its generalizability, is then applied to capture complex, unmodeled effects, despite inherent delay and transient error during adaptation. Formal Lyapunov analysis with explicit stability bounds guarantees convergence under bounded learning errors. The structured design confines learning to the simplest part of the dynamics, enhancing data efficiency while preserving physical interpretability. The approach is generic, with a free-floating space manipulator orientation control task, including integrated null-space collision avoidance, serving as a case study to demonstrate robust tracking performance and applicability to broader robotic domains.
\end{abstract}


\section{INTRODUCTION}
Modern robotic systems increasingly operate in complex, uncertain environments, from collaborative human–robot teams on Earth to autonomous manipulators performing On-Orbit Servicing, Assembly, and Manufacturing (OOS-AM) \cite{ellery2019tutorial,papadopoulos2021robotic,tsiotras2023spacecraft,alizadeh2024comprehensive}. This reality demands control architectures that deliver both high performance and safety. Purely model-based controllers can guarantee stability but often struggle with unmodeled effects, while purely data-driven methods like deep learning can handle complexity but typically lack performance, formal guarantees and require vast amounts of data. A middle path is to embed known physics into the learning problem. The port-Hamiltonian (pH) framework is an energy-based modeling paradigm that explicitly separates conservative (energy-storing) dynamics from dissipative and input channels \cite{van2006port,duindam2009modeling}. By preserving fundamental conservation laws and passivity, pH models have proven effective in robotics and control, with recent works beginning to integrate learning-based components \cite{duong2024port,sprangers2015, nageshrao2014}. 

Employing port-Hamiltonian modeling framework for interacting complex environment is not new \cite{secchi2007control,angerer2017port}, however this work provides an alternative viewpoint by introducing a new physical observable—a virtual port—that connects the conservative part of a robot system to all external interactions, including dissipation and energy inputs. This port state acts as the sole channel for power flow between the system's internal energy-preserving dynamics and its environment. It decouples the system into two parts: a Left-Hand Side (LHS) with a precise, known structure derived from the Hamiltonian, subject only to parametric uncertainties (e.g., mass, inertia); and a Right-Hand Side (RHS) that forms an algebraic relation with this new observable, encompassing all exogenous inputs, disturbances, dissipation, and actuator dynamics \cite{eslami2024data,eslami2023sequential}. A complete robotic system involves the interacting Hamiltonians of its mechanical, electrical, and environmental subsystems. The power of this pH decomposition is that these complex interactions can be modeled exclusively through the port observable in the RHS relations, leaving the conservative mechanical Hamiltonian on the LHS unaffected. This principled decomposition allows us to apply a robust, model-based controller to the LHS to ensure stability, while reserving learning-based adaptation for the simpler RHS mapping. Compared to end-to-end learning, this structured approach, which we term Data-Assisted Control in a port-Hamiltonian framework (DAC-pH) \cite{eslami2025generalization}, imposes physical consistency, thereby improving data efficiency and interpretability.

In space robotics, for example, a free-floating space manipulator, a satellite base with an attached robotic arm operating in microgravity, presents a canonical challenge. Without active base thrusters, any motion of the arm causes reaction forces that move the base. Physically, the total linear and angular momentum of the free-floating system is conserved \cite{papadopoulos1993dynamic, xu1991dynamic}. As a result, the manipulator’s motion induces coupled translation and rotation of the satellite \cite{mittal1994nonlinear}. Seminal work showed that this coupling renders the angular-momentum conservation law a non-integrable velocity constraint, making the system mathematically nonholonomic \cite{papadopoulos1993nonholonomic, spong2006robot}. The physical consequence is profound: the spacecraft’s final attitude depends on the entire path of the arm, not just its final configuration. Such path-dependence gives rise to dynamic singularities \cite{nanos2012cartesian, reinhart2017hybrid}. This behavior makes classical fixed-base control algorithms inapplicable without modification. A breakthrough tool, the Generalized Jacobian Matrix (GJM), was developed to manage this complexity by linearly mapping joint velocities to end-effector motion while implicitly embedding the momentum coupling into a single, configuration-dependent matrix \cite{umetani1989resolved}. Following the same strategy and deriving the Hamiltonian from the Lagrangian leads to a structured port-Hamiltonian model, where the new port state is the virtual power-conjugate to the arm’s momentum.

This hybrid philosophy has several advantages. First, it leverages strong physical priors (the Hamiltonian) on the core dynamics, preserving passivity and allowing for Lyapunov-based stability proofs. Second, it isolates nonlinear or uncertain effects (friction, actuator dynamics) for a data-driven component but shields the main control loop from learning errors. Classical adaptive controllers are typically designed for parametric uncertainties \cite{wang2016adaptive}, but struggle with such structural uncertainties. In practice, because data-driven adaptation can be slow or produce transient errors, we introduce an intermediate high-gain decentralized integrator on the RHS. This creates strict time-scale separation: the LHS controller operates much faster, so that bounded errors in the RHS learning do not destabilize the primary loop. A formal Lyapunov analysis confirms that, under this design, the closed-loop system remains globally stable with explicit bounds on learning error. Furthermore, necessary and sufficient conditions for the stability of these decentralized controllers can be immediately checked and verified, even considering computational delays \cite{eslami2016integrity}.

The proposed approach of learning the RHS mapping is well-aligned with recent advances in hybrid modeling that combine known physical structure with neural approximation of complex components, such as Physics-Informed Neural Networks (PINNs) \cite{karniadakis2021physics, raissi2019physics, beucler2021enforcing}. By explicitly modeling the input and dissipation matrices as outputs of neural subnetworks and embedding known structural constraints (e.g., positive definite dissipation via softplus activations), the architecture benefits from interpretability and prior knowledge, thereby improving generalization \cite{lutter2019deep, greydanus2019hamiltonian}. However, several challenges remain. Real-time online learning imposes strict constraints on numerical stability, requiring careful tuning of optimizers to avoid divergence \cite{chen2020physics}. Furthermore, while the hybrid structure aids interpretability, the learned matrices must be continuously monitored to prevent overfitting or loss of physical meaning. Despite these challenges, the modular and physics-informed design represents a highly promising direction for integrating machine learning into DAC.

To demonstrate the approach, we apply it to the challenging problem of orientation control of a free-floating space manipulator’s end-effector with integrated collision avoidance \cite{blaise2023}. We derive the reduced-order pH model of the system by enforcing conservation of momentum. A robust sliding-mode controller \cite{slotine1987on} is designed on the LHS Hamiltonian dynamics to track a desired end-effector attitude. An artificial potential field (APF) \cite{khatib1985real} is projected into the manipulator’s null space to repulse the arm from the satellite body, ensuring joint limits and self-collisions are avoided without affecting the primary orientation task. Meanwhile, all RHS uncertainties are handled by decentralized high-gain integrators that enforce the learned port-to-input mapping. High-fidelity simulations validate that the proposed controller achieves precise attitude tracking while simultaneously steering clear of obstacles.

The core contribution of this work is the formulation of a data-assisted control architecture that utilizes the physically meaningful decomposition of port-Hamiltonian systems to create a principled, interpretable, and data-efficient hybrid control structure, introducing the port variable as a new physical observable for interaction dynamics. This hybrid controller pairs a robust sliding mode law for the well-modeled conservative dynamics (LHS) with a learning-based mapping for the RHS, with a rigorous Lyapunov-based analysis showing that the closed-loop system is stable despite bounded learning errors. Explicit gain conditions ensure large robustness margins by establishing strict time-scale separation, where RHS error modes decay orders of magnitude faster than the LHS dynamics. The derivation and verification of the complete free-floating manipulator model under momentum conservation, with comprehensive simulations demonstrating precise end-effector orientation tracking while simultaneously enforcing joint limits and obstacle avoidance, validates the framework.

The rest of the paper is organized as follows. Section \ref{sec:pH} derives port-Hamiltonian generic decomposed form. Section \ref{sec:sr} presents the Data-Assisted Control design: the reduced-order free-floating manipulator dynamics and casts them in pH, the model-based LHS controller, the RHS learning layer, and the integration of the null-space potential. Then sub-section \ref{ssec:stability} provides the formal stability and robustness analysis. Section \ref{sec:sim} shows the simulation setup and results with validations. Section \ref{sec:conclusion} concludes and outlines future work.

\section{PORT-HAMILTONIAN DECOMPOSITION}\label{sec:pH}

The pH framework is selected as the foundation for this control architecture due to its intrinsic ability to model complex physical systems based on energy principles \cite{secchi2007control,angerer2017port}. This energy-centric view provides a powerful and physically meaningful way to decompose system dynamics, which is essential for synergizing model-based and data-driven control methods. Within this framework, we introduce a new physical observable, the virtual port $\vect{\Pi}$, which acts as the interface between the system's internal, conservative dynamics and all external energy sources and sinks.

Under Hamiltonian mechanics, this port is a virtual force and moment vector acting on the system's momentum dynamics. It represents the sole channel through which power can flow between two fundamentally different parts of the system. This concept allows for a principled decoupling of the dynamics into a LHS, which possesses a precise structure derived from first principles and is subject only to parametric uncertainties (e.g., misidentified mass or inertia), and a RHS, which forms an algebraic relation with this new observable. The RHS encapsulates all exogenous inputs, including control actions, dissipative effects, external disturbances, and even observation errors. A complete robotic system involves the interaction of multiple Hamiltonians—mechanical, electrical, and environmental. The power of the pH decomposition is its ability to model all such interactions through the port observable within the RHS relations, leaving the core mechanical Hamiltonian on the LHS structurally invariant under certain interactions. This isolates the complex, hard-to-model, and often empirical dynamics to a simpler algebraic mapping, making it an ideal target for data-driven methods. In this work, the problem of interaction is studied from a single robot's perspective, where all external interactions are modeled as disturbances within the RHS. Interactions that would impose new kinematic constraints, thereby altering the LHS Hamiltonian itself, are considered outside the current scope.

This concept of a port as a new observable finds a compelling parallel in quantum mechanics. There, the Hamiltonian governs the unitary evolution of a closed system, but interaction with an environment induces non-unitary dynamics. Observables such as position and momentum are non-commuting projections of the underlying quantum state. In the classical limit, these dynamical variables become commuting observables, permitting distinct analysis for each. The port variable $\vect{\Pi}$ is therefore introduced here as a new classical observable that cleanly separates the internal, energy-preserving dynamics from all external interactions. Our control strategy is thus to regulate this observable directly using model-based knowledge on the LHS, and then use the algebraic relation on the RHS to find the best control input.

\subsection{System Formulation}
A general port-Hamiltonian system is described by the state-space representation:
\begin{align}
    \dot{\vect{x}} = (\vect{J}(\vect{x}) - \vect{R}(\vect{x}))\nabla\mathcal{H}(\vect{x}) + \vect{g}(\vect{x}){u}
\end{align}
Here, the system state vector $\vect{x} \in \mathbb{R}^{2n}$ is composed of $n$ generalized coordinates and $n$ generalized momenta. The total stored energy of the system is given by the scalar Hamiltonian function $\mathcal{H}(\vect{x})$, and its gradient is denoted by $\nabla\mathcal{H}(\vect{x}) \in \mathbb{R}^{2n}$. The matrix $\vect{J}(\vect{x}) \in \mathbb{R}^{2n \times 2n}$ is a skew-symmetric ($\vect{J} = -\vect{J}^T$) interconnection matrix that governs the internal, power-preserving energy exchange. The matrix $\vect{R}(\vect{x}) \in \mathbb{R}^{2n \times 2n}$ is a symmetric, positive semi-definite dissipation matrix modeling energy loss. Finally, the input matrix $\vect{g}(\vect{x}) \in \mathbb{R}^{2n \times m}$ maps the vector of $m$ control inputs ${u} \in \mathbb{R}^m$ to the state dynamics.

For mechanical systems derived from a Lagrangian, the state vector is explicitly defined by the generalized coordinates $\vect{q} \in \mathbb{R}^n$ (e.g., joint angles) and the generalized momenta $\vect{p} \in \mathbb{R}^n$, such that $\vect{x} = \begin{bmatrix} \vect{q}^T & \vect{p}^T \end{bmatrix}^T$. The interconnection matrix $\vect{J}(\vect{x})$ takes the canonical form:
\begin{align}
    \vect{J}(\vect{x}) = \begin{bmatrix} \emptyset & \mathbb{I} \\ -\mathbb{I} & \emptyset \end{bmatrix}
\end{align}
where $\mathbb{I} \in \mathbb{R}^{n \times n}$ is the identity matrix and $\emptyset \in \mathbb{R}^{n \times n}$ is the zero matrix.

\subsection{The LHS-RHS Decomposition}
To formalize the decomposition, we rearrange the standard pH equation to isolate the conservative and non-conservative terms. This introduces the port variable $\vect{\Pi} \in \mathbb{R}^{2n}$, which represents the net generalized force and moment vector resulting from all non-conservative effects:
\begin{align}\label{equ:gen_pH}
    \dot{\vect{x}} - \vect{J}(\vect{x})\nabla\mathcal{H}(\vect{x}) = \vect{\Pi} = -\vect{R}(\vect{x})\nabla\mathcal{H}(\vect{x}) + \vect{g}(\vect{x}){u}
\end{align}
This equation naturally defines the DAC-pH split. The LHS represents the pure, energy-preserving dynamics derived from Hamiltonian mechanics. The RHS is an algebraic expression that defines the port variable $\vect{\Pi}$ as a function of the system state, dissipation, and control inputs. The RHS is where most structural uncertainty appears. Using the notation $(\tilde{\cdot})$ for uncertain parts, we assume the true RHS is given by:
\begin{align}
    \vect{\Pi} = -\tilde{\vect{R}}(\vect{x}) \nabla\mathcal{H}(\vect{x}) + \tilde{\vect{g}}(\vect{x}){u}
\end{align}
For mechanical systems, dissipation and external inputs typically act on the momentum dynamics, not directly on the configuration variables. This imparts a specific block structure to the port variable $\vect{\Pi}$ and the uncertain matrices. The port variable becomes $\vect{\Pi} = \begin{bmatrix} \emptyset & \vect{\tau}^T \end{bmatrix}^T$, where $\vect{\tau} \in \mathbb{R}^n$ is the vector of generalized forces/torques. Consequently, the uncertain dissipation and input matrices are written as:
\begin{align}
    \tilde{\vect{R}}(\vect{x}) = \begin{bmatrix} \emptyset & \emptyset\\ \emptyset & \tilde{\vect{D}}(\vect{x}) \end{bmatrix}, \quad \text{and}\quad \tilde{\vect{g}}(\vect{x}) = \begin{bmatrix} \emptyset & \emptyset\\ \tilde{\vect{B}}(\vect{x}) & \mathbb{I} \end{bmatrix}
\end{align}
where $\tilde{\vect{D}}(\vect{x}) \in \mathbb{R}^{n \times n}$ and $\tilde{\vect{B}}(\vect{x}) \in \mathbb{R}^{n \times m}$ are the uncertain dissipation and control input matrices associated with the momentum dynamics, respectively. All external interactions at the current work such as external moments acting on the joints are included in disturbance input $d(t)$ from other ineracting pH  systems, and generalized input is defined as ${u}=\begin{bmatrix}
    u(t)^T & d(t)^T\\
\end{bmatrix}^T$ where $u$ is the control input from electrical system, e.g. electrical torque. Substituting these into Eq. \eqref{equ:gen_pH} and extracting the second block of equations (corresponding to the momentum dynamics) yields the final decomposed form used for control design:
\begin{align}\label{eq:decomposed}
    \underbrace{\dot{\vect{p}} + \frac{\partial \mathcal{H}}{\partial {\vect{q}}} =  }_{\text{LHS}}\vect{\tau} \overbrace{ = -\tilde{\vect{D}}(\vect{x})\frac{\partial \mathcal{H}}{\partial {\vect{p}}}+\tilde{\vect{B}}(\vect{x})u + d(t)}^{\text{RHS}}
\end{align}
This equation is the cornerstone of our approach. The LHS contains the well-modeled, conservative dynamics, for which a robust controller can be designed to compute a desired torque $\vect{\tau}_c$. The RHS provides a direct algebraic relationship that must be satisfied to generate this torque, making it an ideal target for a learning-based component that maps the desired $\vect{\tau}_c$ to the required physical control input $u$.

\section{SPACE-ROBOT ATTITUDE CONTROL}\label{sec:sr}
The foundational step in designing a controller for any physical system is the development of a mathematical model that accurately describes its dynamic behavior. This is particularly crucial for a free-floating space robot, which constitutes a complex multibody system operating in a zero-gravity environment. The primary challenge lies in capturing the strong dynamic coupling between the manipulator's motion and the unrestricted motion of the spacecraft base. This coupling, governed by the conservation of momentum, introduces significant complexity into the derivation of the equations of motion, which must be systematically addressed to formulate the system in the port-Hamiltonian framework.

\subsection{Modeling}\label{ssec:modeling}
The modeling process begins with the system's kinematics to describe its motion, followed by the application of conservation laws to derive a reduced-order dynamic model suitable for control design.

\subsubsection{Kinematics}
We establish an inertial reference frame, $O_I$, and attach a body-fixed frame, $O_B$, to the center of mass of the spacecraft base (body 0). The full configuration of the system is described by the position and orientation of the base, $\vect{q}_0 \in SE(3)$, and the manipulator's $n$ joint angles, $\vect{q} \in \mathbb{R}^n$. The generalized velocity of the base is $\dot{\vect{q}}_0 =\begin{bmatrix} \dot{\vect{r}}_0^T & \vect{\omega}_0^T \end{bmatrix}^T \in \mathbb{R}^6$, where $\dot{\vect{r}}_0$ and $\vect{\omega}_0$ are the linear and angular velocities of the base frame, respectively, with respect to the inertial frame. The vector of joint velocities is $\dot{\vect{q}} \in \mathbb{R}^n$.

The absolute velocity of the center of mass of the $k$-th link is given by the standard velocity composition rule. The linear velocity $\dot{\vect{r}}_k \in \mathbb{R}^3$ and angular velocity $\vect{\omega}_k \in \mathbb{R}^3$ can be expressed compactly in matrix form. Using the skew-symmetric matrix $S(\cdot)$ to represent the cross-product operation, the spatial velocity of the $k$-th link, $\vect{v}_k =\begin{bmatrix} \dot{\vect{r}}_k^T & \vect{\omega}_k^T \end{bmatrix}^T \in \mathbb{R}^6$, is:
\begin{equation}
   \vect{v}_k = \vect{H}_k \dot{\vect{q}}_0 + \vect{J}_k \dot{\vect{q}}
\end{equation}

\subsubsection{Conservation Laws and the Generalized Jacobian Matrix}
For a free-floating robot with no external forces or torques, the total linear momentum, $\vect{h}_L \in \mathbb{R}^3$, and angular momentum, $\vect{h}_A \in \mathbb{R}^3$, are conserved. Assuming the system starts from rest, they remain zero:
\begin{align}
    &\vect{h}_L = \sum_{k=0}^{N} m_k \dot{\vect{r}}_k = \emptyset, \nonumber\\
    &\vect{h}_A = \sum_{k=0}^{N} (\vect{I}_k \vect{\omega}_k + m_k \vect{r}_k \times \dot{\vect{r}}_k) = \emptyset
\end{align}
where $m_k$ and $\vect{I}_k$ are the mass and inertia tensor of the $k$-th body. Substituting the kinematic relationships, these conservation laws can be expressed as a single matrix constraint on the system velocities:
\begin{equation}
    \vect{H}_{q_0} \dot{\vect{q}}_0 + \vect{H}_q \dot{\vect{q}} = \emptyset
\end{equation}
where the matrices $\vect{H}_{q_0} \in \mathbb{R}^{6 \times 6}$ and $\vect{H}_q \in \mathbb{R}^{6 \times n}$ are functions of the system's mass properties and configuration. This equation represents a non-integrable velocity constraint due to the angular momentum component, rendering the system nonholonomic \cite{papadopoulos1993nonholonomic}. The base velocity can be solved in terms of the joint velocities:
\begin{equation}
    \dot{\vect{q}}_0 = -\vect{H}_{q_0}^{-1} \vect{H}_q \dot{\vect{q}}
\end{equation}
Substituting this back into the link velocity equation eliminates the explicit dependency on the base velocity, resulting in a direct mapping from the joint velocities $\dot{\vect{q}}$ to the absolute link velocities:
\begin{equation}
    \vect{v}_k = \left( \vect{J}_k - \vect{H}_k \vect{H}_{q_0}^{-1} \vect{H}_q \right) \dot{\vect{q}} = \bar{\vect{J}}_k \dot{\vect{q}}
\end{equation}
The matrix $\bar{\vect{J}}_k \in \mathbb{R}^{6 \times n}$ is the renowned Generalized Jacobian Matrix (GJM) \cite{umetani1989resolved}. It implicitly embeds the complex momentum conservation laws, relating the controlled joint velocities directly to the inertial motion of the links.

\subsubsection{Reduced-Order Dynamics and Port-Hamiltonian Formulation}
Using the GJM, we can derive a reduced-order dynamic model in terms of the manipulator joint coordinates $\vect{q}$. In a zero-gravity environment, the system's Lagrangian is equal to its total kinetic energy, $K$. Expressing the link velocities using the GJM, the kinetic energy becomes a quadratic form of the joint velocities:
\begin{align}
     K(\vect{q},\dot{\vect{q}}) & = \frac{1}{2} \dot{\vect{q}}^T \left( \sum_{k=0}^{N} m_k \bar{\vect{J}}_{vk}^T \bar{\vect{J}}_{vk} + \bar{\vect{J}}_{\omega k}^T \vect{I}_k 
    \bar{\vect{J}}_{\omega k} \right) \dot{\vect{q}} \nonumber\\
    & = \frac{1}{2} \dot{\vect{q}}^T M(\vect{q}) \dot{\vect{q}}
\end{align}
This defines the symmetric, positive-definite reduced-order mass matrix $M(\vect{q}) \in \mathbb{R}^{n \times n}$. The Euler-Lagrange equations of motion for the manipulator joints are then given by the standard form:
\begin{equation}
    M(\vect{q})\ddot{\vect{q}} + C(\vect{q},\dot{\vect{q}})\dot{\vect{q}} = \vect{\tau}
\end{equation}
where $C(\vect{q},\dot{\vect{q}}) \in \mathbb{R}^{n \times n}$ is the matrix of Coriolis and centrifugal terms, and $\vect{\tau} \in \mathbb{R}^n$ is the vector of applied joint torques. A key property for control design is that the matrix $(\dot{M} - 2C)$ is skew-symmetric.

To cast this system into the port-Hamiltonian framework, we define the Hamiltonian $\mathcal{H}$ as the total kinetic energy and the generalized momenta as $\vect{p} = M(\vect{q})\dot{\vect{q}}$. The Hamiltonian can then be written purely in terms of state variables $\vect{q}$ and $\vect{p}$:
\begin{align}
    \mathcal{H}(\vect{q},\vect{p}) = \dfrac{1}{2}\vect{p}^TM^{-1}(\vect{q})\vect{p}
\end{align}
From this definition, it follows that $\partial\mathcal{H}/\partial\vect{p} = M^{-1}(\vect{q})\vect{p} = \dot{\vect{q}}$. The dynamics are governed by Hamilton's equations:
\begin{align}\label{equ:hamilt}
    \dot{\vect{q}} = \dfrac{\partial\mathcal{H}}{\partial \vect{p}} \quad \text{and}\quad \dot{\vect{p}} = -\dfrac{\partial\mathcal{H}}{\partial \vect{q}}+\vect{\tau}
\end{align}
To find an expression for $\partial\mathcal{H}/\partial\vect{q}$, we compare the Hamiltonian and Lagrangian dynamics. Taking the time derivative of the momentum definition gives $\dot{\vect{p}} = \dot{M}(\vect{q})\dot{\vect{q}} + M(\vect{q})\ddot{\vect{q}}$. Substituting the Lagrangian equation of motion, $M(\vect{q})\ddot{\vect{q}} = \vect{\tau} - C(\vect{q},\dot{\vect{q}})\dot{\vect{q}}$, and replacing $C(\vect{q},\dot{\vect{q}})=C'(\vect{q},\vect{p})$ yields:
\begin{equation}
    \dot{\vect{p}} = \dot{M}(\vect{q})\dot{\vect{q}} + (\vect{\tau} - C'(\vect{q},\vect{p})\dot{\vect{q}}) = \left(\dot{M}(\vect{q}) - C'(\vect{q},\vect{p})\right)\dot{\vect{q}} + \vect{\tau}
\end{equation}
Comparing this with Hamilton's equation for $\dot{\vect{p}}$ in \eqref{equ:hamilt}, we identify the gradient with respect to configuration:
\begin{align}
    \frac{\partial \mathcal{H}}{\partial {\vect{q}}} = -\left(\dot{M}(\vect{q}) - C'(\vect{q},\vect{p})\right){M}^{-1}(\vect{q})\vect{p}
\end{align}
Finally, Hamilton's equation for the momentum dynamics can be rearranged to perfectly match the decomposed structure introduced in Section \ref{sec:pH}:
\begin{align}
    \underbrace{\dot{\vect{p}} + \frac{\partial \mathcal{H}}{\partial {\vect{q}}}}_{\text{LHS}} = \vect{\tau}
\end{align}
This result is central to our control design. The entire complex, coupled, and nonholonomic dynamics of the free-floating space manipulator are encapsulated in the conservative LHS of the port-Hamiltonian system. The vector of joint torques $\vect{\tau}$ is precisely the port variable that mediates all energy exchange with this conservative core. The RHS, which relates this port to the physical actuator inputs ${u}$ and dissipative effects, is given by:
\begin{align}
    \vect{\tau} = \overbrace{-\tilde{\vect{D}}(\vect{x})\dot{\vect{q}}+\tilde{\vect{B}}(\vect{x})u + d(t)}^{\text{RHS}}
\end{align}
This completes the modeling and decomposition, providing a clear and physically meaningful structure for the application of the DAC framework.

\subsection{LHS Model-Based Control}\label{ssec:lhs}
The control design for the LHS is predicated on a dual-objective strategy. The primary objective is to ensure the end-effector's orientation, $\alpha \in \mathbb{R}$, accurately tracks a desired time-varying reference trajectory, $\alpha_d(t)$. The secondary objective is to leverage the system's redundancy to perform collision avoidance, preventing the manipulator from colliding with its own base.

\subsubsection{Task-Space Kinematics and Redundancy}
To control the orientation, we first establish the kinematic relationship between its velocity, $\dot{\alpha}$, and the joint velocities, $\dot{\vect{q}}$. The end-effector's spatial velocity is given by $\vect{v}_n = \bar{\vect{J}}_n(\vect{q})\dot{\vect{q}}$. For the planar case, the angular velocity component of $\vect{v}_n$ is simply $\dot{\alpha}$. We can therefore extract the corresponding row from the GJM to define a task-space Jacobian, $\vect{D}(\vect{q}) \in \mathbb{R}^{1 \times n}$:
\begin{equation}
    \dot{\alpha} = \vect{D}(\vect{q})\dot{\vect{q}}
\end{equation}
Since the manipulator has more joints than are required for the task ($n > 1$), the system is redundant. The general solution for the joint velocities that satisfy the primary task can be found using the Moore-Penrose pseudoinverse, $\vect{D}^{\dagger}(\vect{q}) = \vect{D}^T(\vect{D}\vect{D}^T)^{-1}$:
\begin{equation}
    \dot{\vect{q}} = \vect{D}^{\dagger}(\vect{q})\dot{\alpha}_{d} + (\mathbb{I} - \vect{D}^{\dagger}(\vect{q})\vect{D}(\vect{q}))\vect{\xi}
\end{equation}
The first term achieves the primary task, while the second term projects an arbitrary joint velocity vector, $\vect{\xi} \in \mathbb{R}^n$, into the null space of the task. Motion in this null space reconfigures the robot's joints without affecting the end-effector's orientation, and can thus be used for the secondary objective.

\subsubsection{Sliding Mode Control and Collision Avoidance}
A robust control law based on the sliding mode control methodology of Slotine and Li \cite{slotine1987on} is designed to compute the desired port value, $\vect{\tau}_c$. We define a sliding surface vector, $\vect{s} \in \mathbb{R}^n$, in joint space:
\begin{align}
    \vect{s} &= \dot{\vect{q}} - \vect{\nu} = M^{-1}(\vect{q})\vect{p} - \vect{\nu}
\end{align}
where $\vect{\nu} \in \mathbb{R}^n$ is a reference joint velocity. This reference velocity is designed to achieve both control objectives simultaneously:
\begin{align}
    \vect{\nu} &= \vect{D}^{\dagger}(\vect{q})(\dot{\alpha}_d - \Lambda\tilde{\alpha}) + (\mathbb{I} - \vect{D}^{\dagger}(\vect{q})\vect{D}(\vect{q}))\vect{\xi}
\end{align}
where $\tilde{\alpha} = \alpha - \alpha_d$ is the orientation tracking error and $\Lambda \in \mathbb{R}_{++}$ is a positive feedback gain. The null-space velocity, $\vect{\xi}$, is used for collision avoidance via the APF method \cite{khatib1985real}. A potential function, $U(\vect{q}) \in \mathbb{R}$, is designed to be high near obstacles (e.g., the spacecraft base) and low in safe regions. The null-space velocity is then set to move the system "downhill" along the negative gradient of this potential:
\begin{equation}
    \vect{\xi} = -\eta \nabla U(\vect{q})
\end{equation}
where $\eta \in \mathbb{R}_{++}$ is a positive gain. The control law for the desired port torque, $\vect{\tau}_c$, is chosen to drive $\vect{s}$ to zero:
\begin{equation}
    \vect{\tau}_c = M(\vect{q})\dot{\vect{\nu}} + C'(\vect{q},\vect{p})\vect{\nu} - K_d\vect{s}
\end{equation}
where $K_d \in \mathbb{R}^{n \times n}$ is a symmetric, positive-definite gain matrix. Applying this control torque to the LHS of the decomposed system in \eqref{eq:decomposed} yields the closed-loop error dynamics on the sliding surface:
\begin{align}
    M(\vect{q})\dot{\vect{s}} + (C'(\vect{q},\vect{p})+K_d)\vect{s} = \emptyset
\end{align}
This ensures that $\vect{s} \to 0$, which in turn guarantees that the tracking error dynamics converge according to $\dot{\tilde{\alpha}}+\Lambda \tilde{\alpha} = 0$, ensuring $\tilde{\alpha} \to 0$. To formalize this, consider the Lyapunov candidate function $V = \frac{1}{2}\vect{s}^TM(\vect{q})\vect{s}$. The time derivative of $V$ is:
\begin{align}
  \dot{V} = \vect{s}^TM\dot{\vect{s}} + \frac{1}{2}\vect{s}^T\dot{M}\vect{s}
\end{align}
Substituting the error dynamics $M\dot{\vect{s}} = -(C'+K_d)\vect{s}$ gives:
\begin{align}
  \dot{V} &= \vect{s}^T(-(C'+K_d)\vect{s}) + \frac{1}{2}\vect{s}^T\dot{M}\vect{s} \nonumber\\
  & = \frac{1}{2}\vect{s}^T(\dot{M}-2C')\vect{s} - \vect{s}^TK_d\vect{s}
\end{align}

Using the skew-symmetric property of the matrix $(\dot{M}-2C')$, the term $\frac{1}{2}\vect{s}^T(\dot{M}-2C')\vect{s}$ is identically zero. Thus, the derivative simplifies to $\dot{V} = -\vect{s}^TK_d\vect{s}$. Since $K_d$ is positive definite, this guarantees that $\vect{s}$ converges to zero. The bounds on the convergence rate can be established from the inequalities:
\begin{align}
    &\;\;\;\frac{1}{2}\underline{\lambda}(M)\|\vect{s}\|^2 \leq V \leq \frac{1}{2}\overline{\lambda}(M)\|\vect{s}\|^2 \quad \text{and}\nonumber\\& -\overline{\lambda}(K_d)\|\vect{s}\|^2 \leq \dot{V} \leq -\underline{\lambda}(K_d)\|\vect{s}\|^2
\end{align}
Combining these yields a differential inequality for $V$:
\begin{align}
    -\frac{2\overline{\lambda}(K_d)}{\underline{\underline{\lambda}}(M)} V \leq \dot{V} \leq -\frac{2\underline{\lambda}(K_d)}{\overline{\overline{\lambda}}(M)} V
\end{align}
where $\underline{\underline{\lambda}}(M)$ and $\overline{\overline{\lambda}}(M)$ respectively mean infimum and supremum  eigenvalues of $M$ over configuration space $\vect{q}$. Integrating this inequality provides the exponential bounds on the convergence of $V(t)$:
\begin{align}
    V(0)e^{-\dfrac{2\overline{\lambda}(K_d)}{\underline{\underline{\lambda}}(M)}t} \leq V(t) \leq V(0)e^{-\dfrac{2\underline{\lambda}(K_d)}{\overline{\overline{\lambda}}(M)}t}
\end{align}
This confirms the exponential convergence of the sliding surface to zero. The upper bound on $V(t)$ is determined by the ratio of the minimum control gain to the maximum system inertia, while the lower bound is determined by the ratio of the maximum gain to the minimum inertia. These bounds are crucial for the subsequent design of the RHS controller.

\subsection{RHS Control and Learning}\label{ssec:rhs}
Utilizing the precisely known dynamics of the LHS as a high-fidelity observable, and measuring the generalized coordinates and their derivatives ($\breve{\vect{q}}$, $\breve{\dot{\vect{q}}}$, $\breve{\ddot{\vect{q}}}$), we can compute the observed generalized forces and moments at the interaction port, $\breve{\vect{\tau}}$. The superscript $\breve{\cdot}$ denotes an observed, measured, or calculated value.
\begin{align}
    \breve{\vect{\tau}} = M(\breve{\vect{q}})\breve{\ddot{\vect{q}}} + C(\breve{\vect{q}},\breve{\dot{\vect{q}}})\breve{\dot{\vect{q}}}
\end{align}
This observable serves as the ground-truth signal for identifying the RHS mapping, which encapsulates all dissipative, actuation, and external disturbance effects. While the LHS can be subject to parametric uncertainties, this work assumes it is known; standard adaptive methods could be employed to mitigate the effect of such uncertainties. The RHS mapping is modeled as:
\begin{align}\label{equ:rhs_data_model}
    \breve{\vect{\tau}} = \tilde{B}(\vect{x}){u} - \tilde{D}(\vect{x})\dot{\vect{q}} + {d}(t)
\end{align}
The identification of this mapping is a supervised regression problem. However, applying standard data-driven techniques in a real-time control context presents significant challenges, including limitations of classical methods like RLS \cite{ljung1998system}, the basis-dependency of Koopman operator methods \cite{eslami2024data}, and the real-time feasibility and stability concerns of generic DNNs \cite{chen2020physics, greydanus2019hamiltonian}.

These limitations motivate a hierarchical approach that decouples the problem of stability from that of performance enhancement. We propose a framework comprising two main components: (1) an inner-loop controller to adjust the RHS error convergence and rapidly attenuate errors, and (2) a higher-level, physics-informed learning module that identifies the RHS parameters to improve performance. This modular design provides a stable foundation upon which advanced learning algorithms can be safely deployed.

Starting from the upper layer, a modular network similar to Physics-Informed Neural Network is proposed, chosen for its ability to embed physical structure and its generalization capabilities \cite{raissi2019physics, karniadakis2021physics}. The network takes a feature vector $\vect{X}(\vect{x}) \in \mathbb{R}^{p}$, constructed from $p$ features of the state $\vect{x}$ (e.g., products, powers, sinusoids and etc), as input. This feature vector is processed by two separate hidden layer modules to form distinct functional bases for the $\hat{B}$ and $\hat{D}$ matrices, as follows: 
\begin{align*}
    \vect{h}_B(\vect{x}) &=\sigma_B^{1}\left(\ldots\sigma_B^{l_B}({W}^{l_B}_{B} \vect{X}(\vect{x}) + \vect{b}^{l_B}_{B})+\ldots)+\vect{b}^{1}_B\right)\\
    \vect{h}_D(\vect{x}) &=\sigma_D^{1}\left(\ldots\sigma_D^{l_D}({W}^{l_D}_{D} \vect{X}(\vect{x}) + \vect{b}^{l_D}_{D})+\ldots)+\vect{b}^{1}_D\right)
\end{align*}
where $\sigma^{l_B}_B, \sigma^{l_D}_D$ are activation functions for hidden layers $l_B$ and $l_D$ showing the depth of network. These hidden representations are then fed into separate linear output heads:
\begin{align}
    \hat{B}(\vect{x}) &= \text{mat}\left({W}_{B} \vect{h}_B(\vect{x}) + \vect{b}_{B}\right) \\
    \hat{D}(\vect{x}) &= \text{diag}\left(\sigma_{sp}({W}_{D} \vect{h}_D(\vect{x}) + \vect{b}_{D})\right)
\end{align}
where $\sigma_{sp}$ is the softplus activation to enforce the positive definiteness of the dissipation matrix $\hat{D}$, and $\text{mat}(\cdot)$ is an operator for generating a matrix from a vector in the proper arrangement. The complete network predicts the RHS port value as $\hat{\tau} = \hat{B}(\vect{x}){u} - \hat{D}(\vect{x})\dot{\vect{q}} + {d}(t)$, and is trained by minimizing the mean squared error loss $\ell = \|\breve{\vect{\tau}} - \hat{\vect{\tau}}\|^2$. We leave the discussion on the effect of $d(t)$ in learned optimal matrices right after introducing the RHS controller.

Now, a fast RHS control loop should be designed to regulate the error $e_{\tau} = \vect{\tau}_c^r - \vect{\tau}_c^g$, where $\vect{\tau}_c^r$ is requested port signal by the LHS and the $\vect{\tau}_g^r$ is actual generated (observed) port signal by the RHS. To design a controller for this nonlinear system, we linearize the dynamics around a nominal operating trajectory $(\vect{x}_0(t), \dot{\vect{q}}_0(t), {u}_0(t))$. At this equilibrium, the nominal error $e_{\tau_0}$ satisfies the steady-state relationship:
\begin{equation*}
    e_{\tau_0} = \vect{\tau}_{c_0}^r - \left( \hat{B}(\vect{x}_0){u}_0 - \hat{D}(\vect{x}_0)\dot{\vect{q}}_0 \right) 
\end{equation*}
Assuming initially there is no disturbance and our knowledge about LHS observation is perfect, i.e., ${d}(t_0)=0$ and $\breve{e}_{0}(t_0)=0$ therfore $\delta \breve{e} = \breve{e}(t)$ and $\delta d  =d(t)$. Let perturbations around this trajectory be denoted by $\delta(\cdot)$. The linearized perturbation error, $\delta e_\tau$, is found by taking the first-order Taylor expansion of the error and subtracting the equilibrium condition. This yields:
\begin{align}
    \delta e_\tau \approx \underbrace{\delta\vect{\tau}_c^r - \delta \breve{e} - \delta d}_{\delta z_1} - \left( \hat{B}_0 \delta {u} - \hat{D}_0 \delta \dot{\vect{q}} + \hat{J}_0 \delta \vect{x} \right) 
\end{align}
where the the Jacobian matrices are $\hat{B}_0 = \hat{B}(\vect{x}_0)$, $\hat{D}_0 = \hat{D}(\vect{x}_0) = \text{diag}(\hat{d}_1, \ldots, \hat{d}_n)$ for $\hat{d}_i\geq 0$, and $\hat{J}_0 = \partial/\partial \vect{x} \left( \hat{D}(\vect{x})\dot{\vect{q}}_0 - \hat{B}(\vect{x}){u}_0 \right)$ at ${\vect{x}=\vect{x}_0}$, which are computed using the analytical form of the PINN.

We design a feedback controller for the perturbations, $\delta {u}(s) = K(s) \delta e_\tau(s)$. Taking the Laplace transform of the linearized equation gives the closed-loop dynamics:
\begin{align}
    \delta e_\tau(s) = \underbrace{(I+\hat{B}_0 K(s))^{-1}\begin{bmatrix} I & \hat{D}_0 & \hat{J}_0 \end{bmatrix}}_{G(s)} \underbrace{\begin{bmatrix} \delta z_1(s) \\ \delta\dot{\vect{q}}(s) \\ \delta\vect{x}(s) \end{bmatrix}}_{\vect{z}(s)}
\end{align}

This equation provides a framework for analyzing how different exogenous inputs ${z}(s)$ propagate to the tracking error through the sensitivity transfer function $G(s)$. The integral controller $K(s)=K/s$ is designed to have high gain at low frequencies, meaning the term $(I+\hat{B}_0 K/s)^{-1}$ provides significant attenuation for low-frequency components of the input vector ${z}(s)$. This ensures that low-frequency disturbances in $\delta d$ and reference commands $\delta\vect{\tau}_c^r$ are effectively rejected, which is the primary role of integral action in control systems. Conversely, high-frequency components, such as measurement noise in $\delta{e}$ or the unmodeled dynamics captured in ${d}_{HF}(t)$, will pass through $G(s)$ with less attenuation, as it behaves like a high-pass filter. While practical measures such as low-pass filtering can mitigate high-frequency measurement noise, the reference signal $\delta\vect{\tau}_c^r$ is generated by the stable, loop LHS dynamics and is thus inherently smooth and accounted for in the controller's design.

The disturbance ${d}(t)$ can originate from various sources and may contain a wide range of frequency components. It is useful to decompose it into a low-frequency part, ${d}_{LF}(t)$, and a high-frequency part, ${d}_{HF}(t)$, relative to the bandwidth of the RHS controller. The integral action of the inner-loop controller provides high gain at low frequencies, making it highly effective at rejecting static and slowly varying disturbances. Consequently, the controller actively compensates for ${d}_{LF}(t)$, while the unattenuated high-frequency component ${d}_{HF}(t)$ remains as a residual that affects the learning process.

The presence of the uncompensated high-frequency disturbance, $\vect{d}_{HF}(t)$, poses a significant challenge for the learning algorithm. A fundamental principle of system identification is that parameter estimates become biased if unmodeled inputs (disturbances) are correlated with the measured inputs used for model fitting \cite{ljung1998system}. In this context, the PINN minimizes the prediction error without explicit knowledge of $\vect{d}_{HF}(t)$. If this disturbance is correlated with the control action $\vect{u}$ or the velocity $\dot{\vect{q}}$, the learning algorithm will result biased estimate. From a control perspective, this is not necessarily a problem. The goal is to obtain a model that accurately predicts the system's response to control inputs, even if that model implicitly includes the average effect of certain disturbances. Since the inner-loop controller is designed to be robust to parameter uncertainties, learning these effective matrices is acceptable, provided they remain within the bounds for which stability is guaranteed.

The general design goal is to shape $G(s)$ to minimize the error norm, $\|\delta e_\tau\| \leq \|G(s)\|\|\vect{z}(s)\|$, using methodologies like $\mathcal{H}_\infty$ synthesis. However, for a type zero system, a decentralized integral controller $K(s) = K/s$, where $K=\text{diag}(k_1, \dots, k_n)$ can vanish errors. The stability of this linearized loop is determined by the roots of the characteristic equation $\det(sI + \hat{B}_0 K) = 0$. The necessary and sufficient conditions for the existence of stabilizing integral gains are related to Decentralized Integral Controllability (DIC), which requires that $\det(\hat{B}_0) \neq 0$ \cite{eslami2016integrity}. By diagonalizing the linearized input matrix $\hat{B}_0$, such that $\hat{T}^{-1}\hat{B}_0\hat{T} = \text{diag}(\hat{\lambda}_1, \dots, \hat{\lambda}_n)$ (assuming distinct eigenvalues without loss of generality), system's modes can be analyzed. The error dynamics can be seen as a linear combination of modes, each passing through a high-pass filter:
\begin{align}
    G_{ij}(s) = \sum_{p=1}^{n} \hat{T}_{ip}\left(\dfrac{s}{s + k_p\hat{\lambda}_p}\right)\hat{\Gamma}_{pj}
\end{align}
where depending on channel input gain $\hat{\Gamma}_{pj}$ can be either of followings,
\begin{align}
    \hat{\Gamma}_{pj} = \{ \hat{T}^{-1}_{pj},  \;\hat{T}^{-1}_{pj}\hat{d}_j,  \;\sum_{m=1}^{n} \hat{T}^{-1}_{pm}\hat{J}_{0_{mj}}\}
\end{align}
Since the effect of gain uncertainty is multiplicative, all amplitude uncertainty can be collected in a worst-case multiplicative term for each mode $p$, denoted as $\hat{\beta}_p=\max_{i,j,q}(\hat{T}_{ip} \hat{\Gamma}_{pj}^{\{q\}} / ({T}_{ip} {\Gamma}_{pj}^{\{q\}}))$. Derivation of performance bounds can proceed as detailed in the following section.

\subsection{Stability Analysis}\label{ssec:stability}
The presence of a tracking error $e_\tau$ in the RHS loop directly impacts the stability of the entire system. Consider the revised LHS control law,
\begin{align}
    \vect{\tau}^r_c = M(\vect{q})\dot{\vect{\nu}} + C'(\vect{q},\vect{p})\vect{\nu} - K_d\vect{s} - \chi \dfrac{\vect{s}}{\|\vect{s}\|}
\end{align}
where the term $-\chi \vect{s}/\|\vect{s}\|$ is included to compensate the effect of tracking error $e_\tau$ satisfying $\|e_\tau\|\leq \chi$ \cite{eslami2019robust}.

The actual generalized force, $\tau_c$, applied to the LHS dynamics is given by,
\begin{align}\label{eq:chi_tau}
    \tau_c = \tau_c^r - e_\tau = M(\vect{q})\dot{\nu} + C'(\vect{q},\vect{p})\nu - K_d\vect{s} - \chi \dfrac{\vect{s}}{\|\vect{s}\|} - e_\tau
\end{align}
Using the same Lyapunov candidate for the LHS, $V = \frac{1}{2}\vect{s}^T M(q) \vect{s}$, differentiating and using the skew-symmetry property of $\dot{M}-2C'$,
\begin{align}
    \dot{V} & = \vect{s}^T M \dot{\vect{s}} + \frac{1}{2}\vect{s}^T \dot{M} \vect{s} \\
    & = \vect{s}^T(-K_d\vect{s} - \chi\dfrac{\vect{s}}{\|\vect{s}\|} - e_\tau) \\
  \nonumber  & = -\vect{s}^T K_d \vect{s} -\chi   \|\vect{s}\| -\vect{s}^T e_\tau
\end{align}
Using $\|e_\tau\|\leq \chi$ and Cauchy–Schwarz inequality, $-\vect{s}^T e_\tau \leq \chi \|\vect{s}\|$. Therefore, 
\begin{align}
    \dot{V} & \leq -\vect{s}^T K_d \vect{s}
\end{align}

Since $K_d$ is positive-definite, it is ensured that $s \rightarrow 0$. However, the discontinuous nature of this term induces chattering, which is typically mitigated by using a smooth approximation like $\tanh(\vect{s}/\epsilon)$, introducing its performance trade-offs. A significant advantage of the control architecture here is having $\chi$ directly from the RHS loop, while getting the least $\chi$ would benefit this trade-off. 

Our framework addresses this challenge by ensuring that $\|e_\tau\|$ vanishes rapidly through RHS control design. By enforcing a strict time-scale separation between the RHS and LHS loops, the dynamics of $e_\tau$ are made orders of magnitude faster than the dynamics of $\vect{s}$. Consequently, as $\vect{s}$ converges, $e_\tau$ has already decayed to a negligible value, ensuring the term $\vect{s}^T e_\tau$ is insignificant and stability is maintained by the term $\vect{s}^T K_d \vect{s}$.

The crucial design principle is this time-scale separation. The slowest decay rate of the RHS loop, $\underline{\sigma}_{RHS} = \min_p k_p \text{Re}(\hat{\lambda}_p)$ for $p=p_m$, must be significantly faster than the fastest mode of the LHS sliding dynamics, $\overline{\lambda}_{LHS_s}$, by a factor of $T_d \gg 1$:
\begin{align}
    \underline{\sigma}_{RHS} > T_d \overline{\lambda}_{LHS_s} = 2T_d \dfrac{\overline{\lambda}(K_d)}{\underline{\underline{\lambda}}(M)}
\end{align}
This baseline must be robust to uncertainties.
To ensure this uncertainty is compensated for within a target settling time, $T_0 = 5/(T_d \overline{\lambda}_{LHS_s})$, the required decay rate for the uncertain modes, $\underline{\sigma}_n$, must satisfy:
\begin{align}
    & \ln{(\hat{\beta}_{p_m})} - (\underline{\sigma}_n - \underline{\sigma}_{RHS})T_0 \leq 1 \implies \nonumber\\
    & \underline{\sigma}_n \geq \underline{\sigma}_{RHS} + \frac{\ln(\hat{\beta}_{p_m})}{T_0}
\end{align}
Substituting the time-scale separation requirement yields a practical design rule for the integrator gains under uncertainty:
\begin{align}
    \underline{\sigma}_n > T_d \overline{\lambda}_{LHS_s} \left(1 + \dfrac{\ln(\hat{\beta}_{p_m})}{5}\right)
\end{align}
This formally connects the controller gains to the required performance ($T_d$), system properties ($\overline{\lambda}_{LHS_s}$), and estimated model uncertainty ($\hat{\beta}_{p_m}$), ensuring a robustly stable inner loop. Assuming almost similar dissipation and control input behavior for all joints, one can assign $k_p=c/\text{Re}(\hat{\lambda}_p)$ with constant $c$, and hence the following inequality guarantees stability of the closed-loop system, and the update rule will be very simple.
\begin{align} \label{equ:c_min}
    c > 2\;{T_d^2} \;\dfrac{\overline{\lambda}(K_d)}{\underline{\underline{\lambda}}(M)} \left(1 + \dfrac{\ln(\overline{\hat{\beta}})}{5}\right)
\end{align}

Note that due to the time-scale separation, the empirically calculated bound $\chi$ in \eqref{eq:chi_tau} can be kept very small. In practice, the LHS dynamics are designed with time constants on the order of seconds, while the RHS integrators can achieve decay rates on the order of milliseconds. This vast separation provides wide gain and phase margins, creating a stable foundation where even slow or aggressive learning algorithms can operate without inducing instability. The learned parameters are employed in a smooth update rule to change the $k_i$ holding the inequality.

\section{SIMULATIONS and VALIDATIONS}\label{sec:sim}
To validate the proposed control architecture, high-fidelity simulations are conducted. While the theoretical framework is developed for a general $n$-DOF manipulator, the simulations focus on a 2-DOF planar space manipulator without loss of generality. This specific case is sufficiently complex to exhibit the key dynamic coupling and nonholonomic behavior characteristic of free-floating systems. Furthermore, its planar nature allows for clear visualization and straightforward verification of the control objectives, namely precise end-effector orientation tracking and simultaneous collision avoidance. The successful performance in this case study serves as a strong validation of the general applicability of the DAC-pH framework to more complex, higher-dimensional space robotic systems.

The simulations are carried out with a robot parameter set of triples $(m_i, l_i, I_i)$ for mass in \texttt{kg}, length in \texttt{m}, and moment of inertia in \texttt{kg.m$^2$} for $i=0,1,2$, corresponding to the base, link 1, and link 2, respectively: $(2, 0.1225, 0.02)$, $(1, 0.3464, 0.01)$, and $(1, 0.3464, 0.01)$.

The LHS controller gains are selected as $K_d=0.5\mathbb{I}$, $\Lambda = \mathbb{I}$, $\epsilon = 0.2$, and $\eta = 0.1$. The RHS controller needs only one initial parameter $c$, which is calculated for the nominal matrices of the RHS according to \eqref{equ:c_min} and is provided in the plots thereafter over time. The ADAM optimizer with parameters $\beta_1=0.9$, $\beta_2=0.999$, and $\varepsilon=10^{-8}$ is employed for the online learning of the neural network.

In order to test the effectiveness of the learning, initial static and non-state-dependent matrices are selected for a warming-up phase, after which nonlinearities dependent on the state variables are introduced. For the warming-up phase, the initial dissipation matrix ${D}$ and control input matrix ${B}$ are:
$$
{D} = \begin{bmatrix} 0.1 & 0 \\ 0 & 0.1 \end{bmatrix}, \quad {B} = \begin{bmatrix} 1 & 0 \\ 0 & 1 \end{bmatrix}
$$
Subsequently, the true underlying matrices, $\tilde{{D}}$ and $\tilde{{B}}$, are introduced with state-dependent linear and quadratic uncertainty terms to simulate more complex dynamics:
\begin{align}
    \Delta {D}(\vect{p}) = \underbrace{\text{diag}(0.02,0.05)+ \text{diag}(0.4, 0.4)\vect{p}}_{\text{Linear uncertainty}}+\underbrace{
    \text{diag}(-0.05,0.1)\vect{p}\odot \vect{p}}_{\text{Nonlinear uncertainty}}, \quad\text{and}
\end{align}
\begin{align}
    \Delta{{B}}(\vect{x}) = \underbrace{ \begin{bmatrix}
        -0.1 & 0\\
        0.05 & 0.15\\
    \end{bmatrix} \vect{q}}_{\text{Linear uncertainty}} + 
    \underbrace{\text{diag}(0.15,- 0.25)\vect{p}\odot \vect{p} + 
    \text{offdiag}(- 0.1,0.4)q_1^2}_{\text{Nonlinear uncertainty}}
\end{align}
where $\odot$ is element-by-element product. These uncertainty terms are introduced to model complex physical phenomena. The ${\Delta D}$ term is designed to capture nonlinear friction behaviors. Meanwhile, nonlinearities in the actuation system, stemming from power changes that depend on the robot's position and velocity as described by the electrical Hamiltonian, are also modeled. The hypothesis network architecture is selected to match this degree of complexity. By using only the state vector $\vect{x}$ as its feature input, the approach demonstrates powerful generalizability compared to traditional methods that require hand-engineered observables. A DNN with two separate hidden layers, i.e., $l_B=2$ and $l_D=2$. In both nets, 16 and 8 neurons for the first and second layers are arranged. As the complexity of the nonlinearities increases, a more sophisticated network with additional hidden layers and feature inputs can be employed, although this requires more powerful real-time computational resources.

\begin{figure}[!ht]
     \centering
     \begin{subfigure}[b]{0.48\textwidth}
         \centering
         \includegraphics[width=\textwidth]{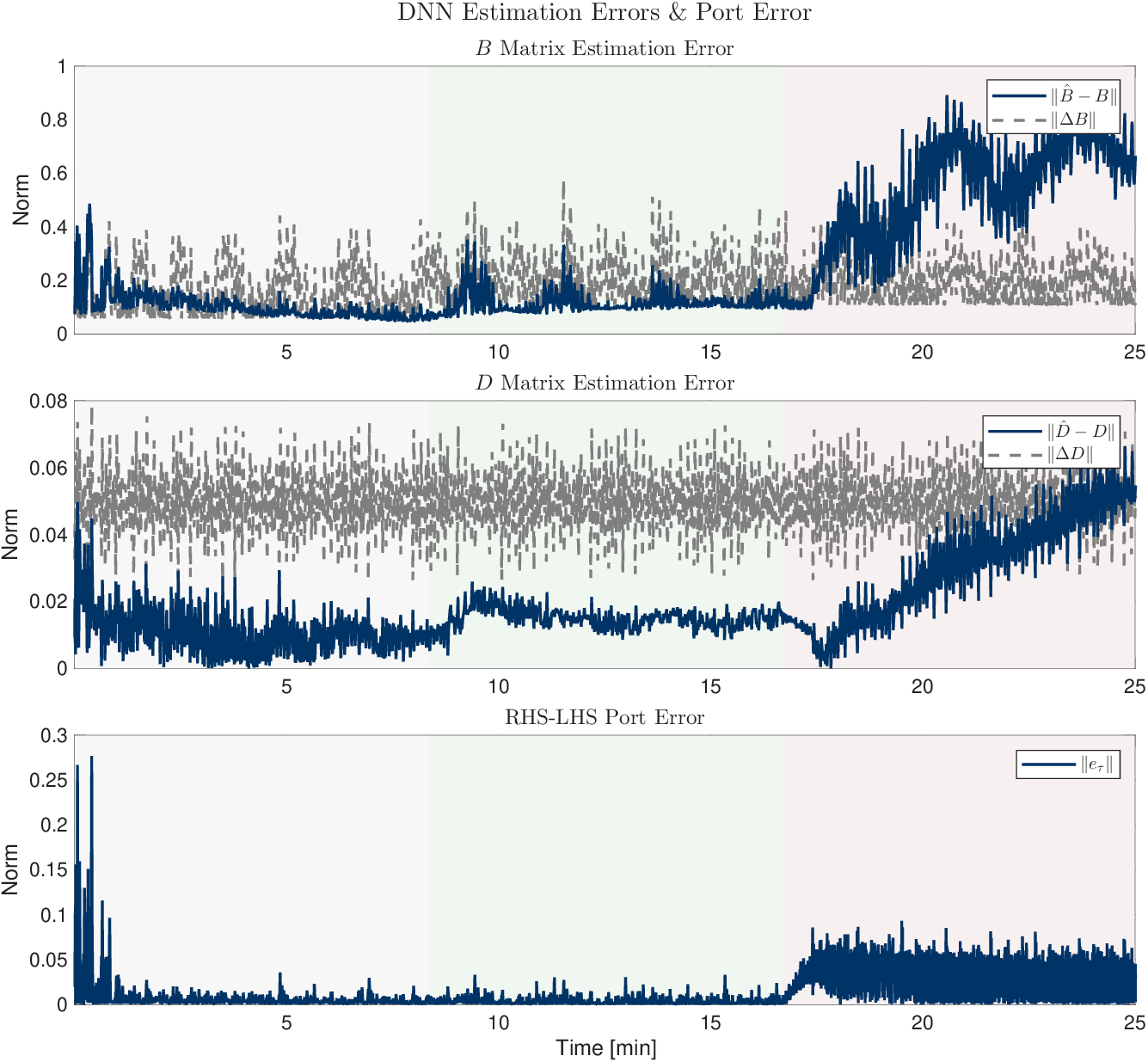}
         \caption{Norm of uncertainty, estimation error, and RHS control error}
         \label{fig:DNN_Performance}
     \end{subfigure}
     \hfill
     \begin{subfigure}[b]{0.48\textwidth}
         \centering
         \includegraphics[width=1\textwidth]{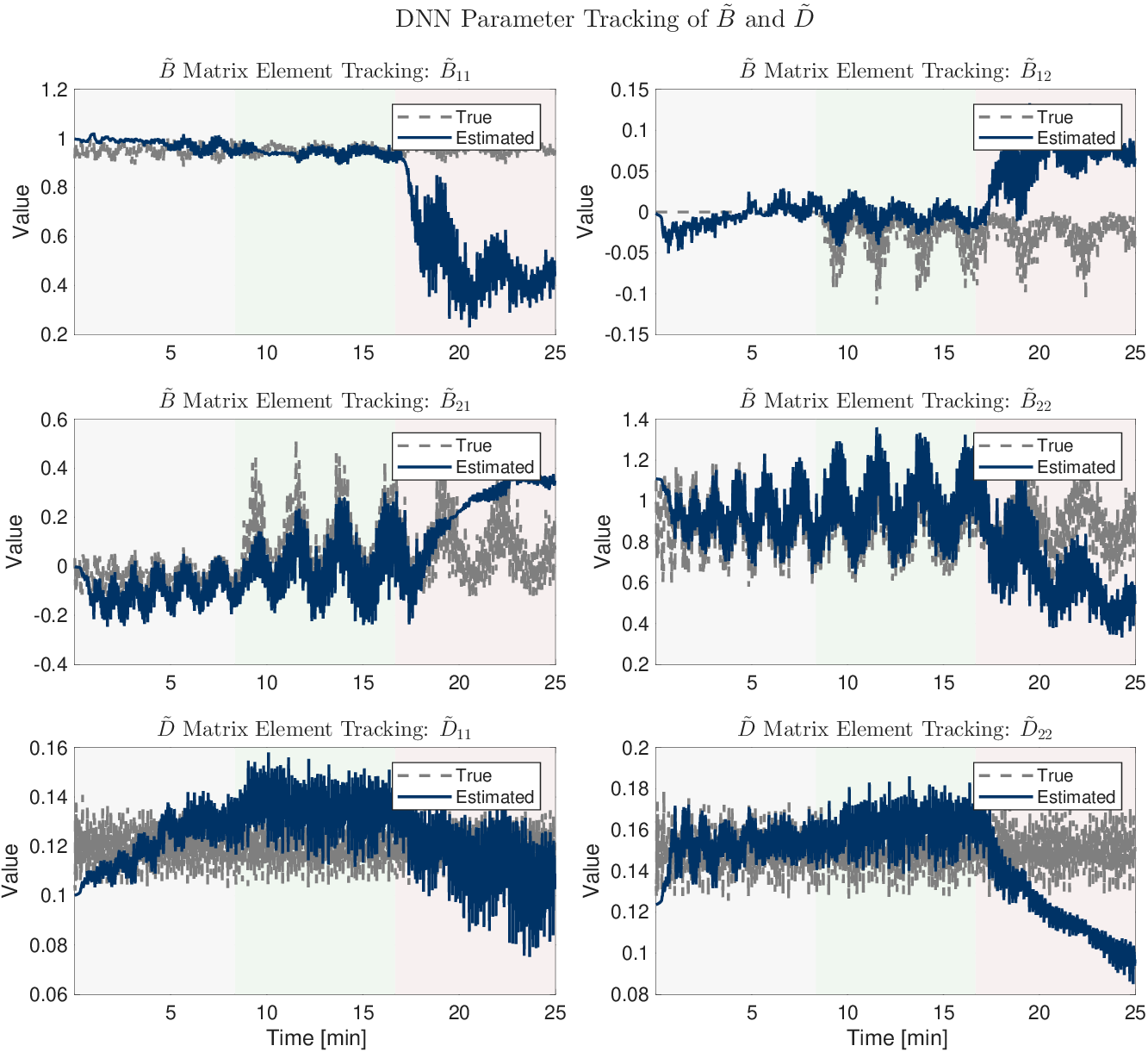}
         \caption{True and estimated entries of $\tilde{B}$ and $\tilde{D}$}
         \label{fig:DNN_Performance_elements}
     \end{subfigure}
        \caption{DNN Performance in identification of uncertain RHS}
        \label{fig:DNN}
\end{figure}

\begin{figure}[!ht]
     \centering
     \begin{subfigure}[b]{0.48\textwidth}
    \centering
    \includegraphics[width=\textwidth]{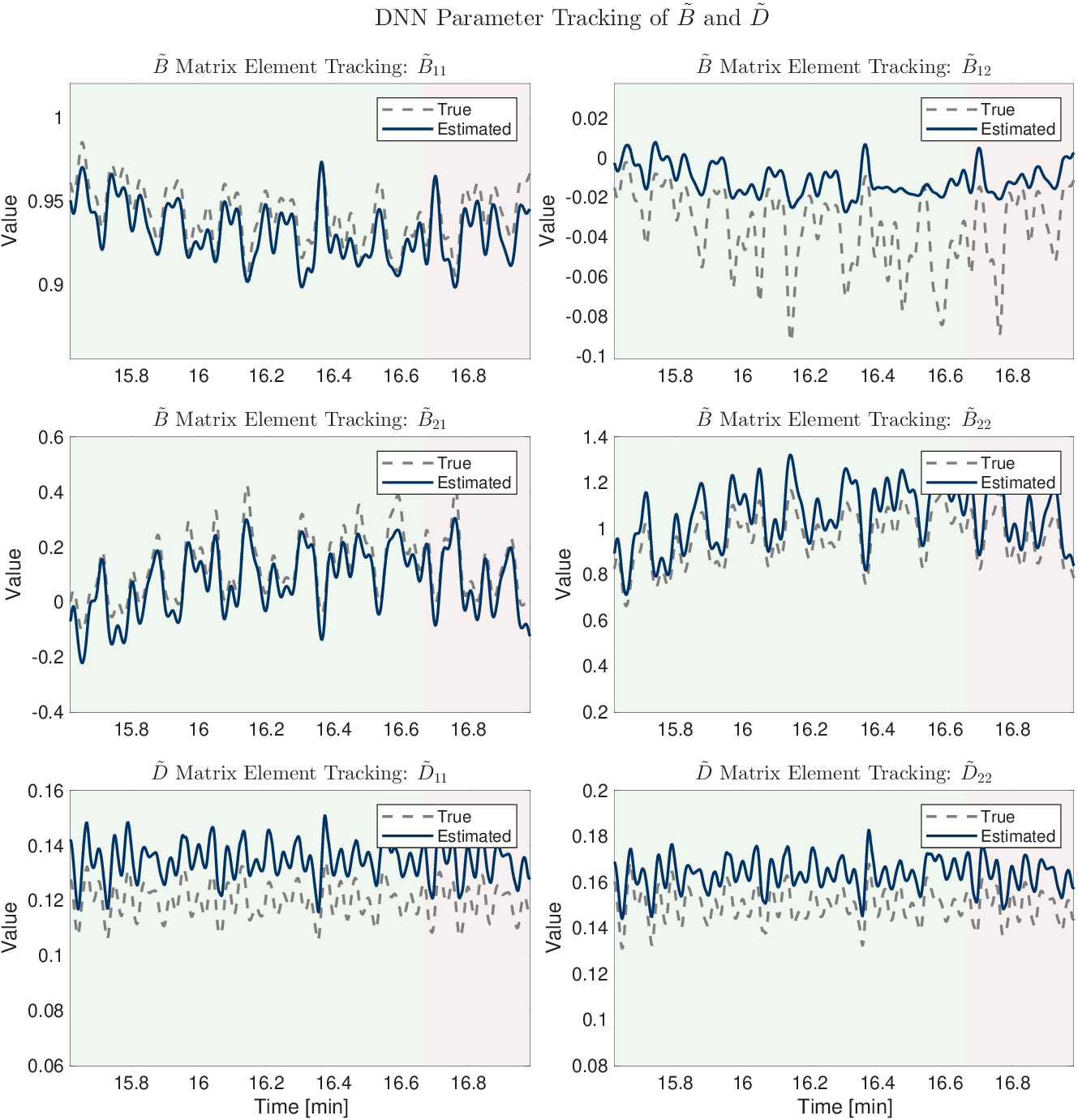}
    \caption{DNN estimation performance at the end of phase 2 of learning}
    \label{fig:DNN_Gen}
     \end{subfigure}
     \hfill
     \begin{subfigure}[b]{0.48\textwidth}
         \centering
         \includegraphics[width=1\textwidth]{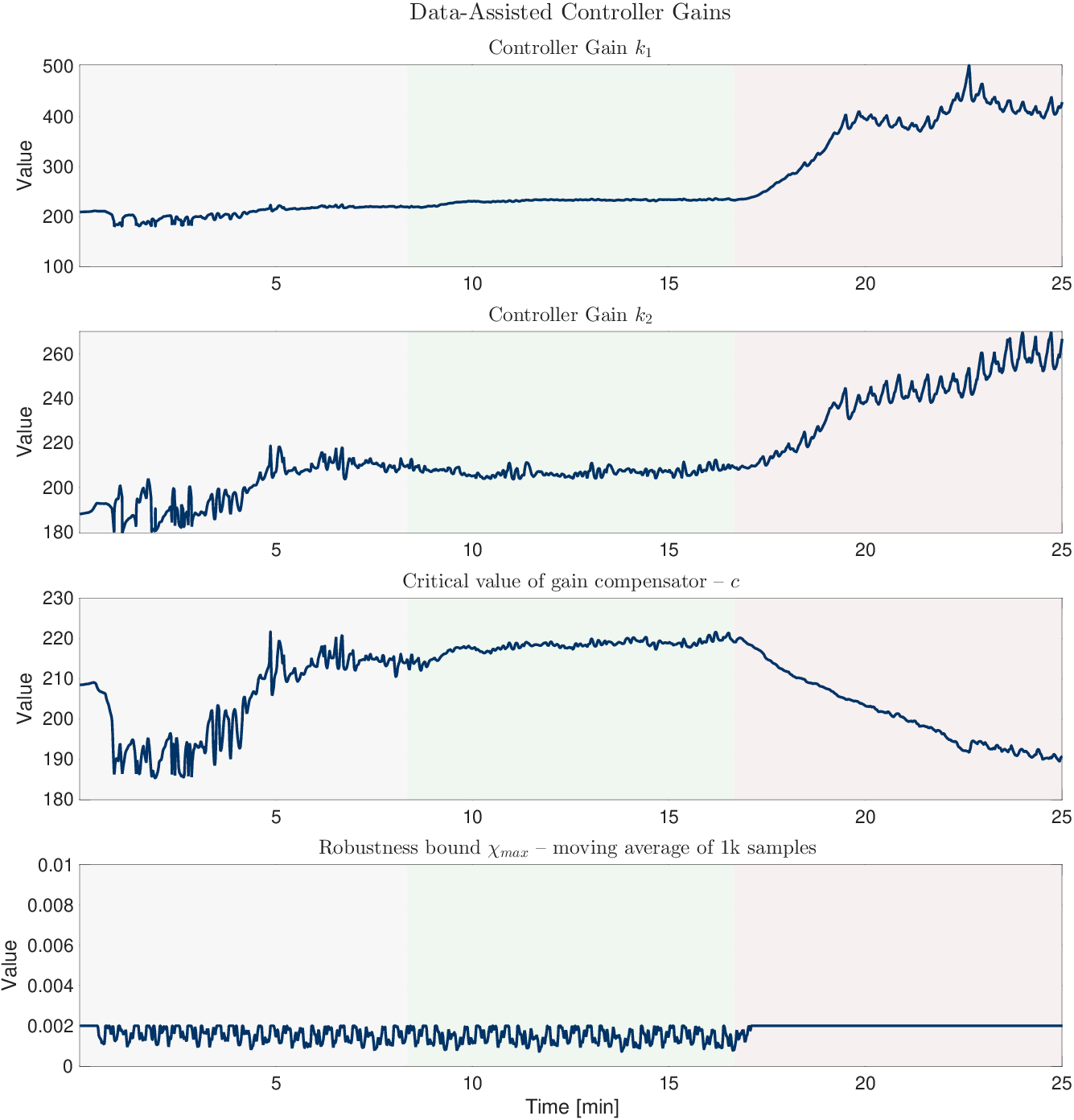}
         \caption{RHS controller gains and LHS uncertainty bound}
         \label{fig:DAC_gains}
     \end{subfigure}
        \caption{Detailed DNN Performance, RHS controller gain variations and LHS uncertainty bound}
        \label{fig:DNN_2}
\end{figure}

\begin{figure}[!ht]
     \centering
     \begin{subfigure}[b]{0.48\textwidth}
         \centering
         \includegraphics[width=\textwidth]{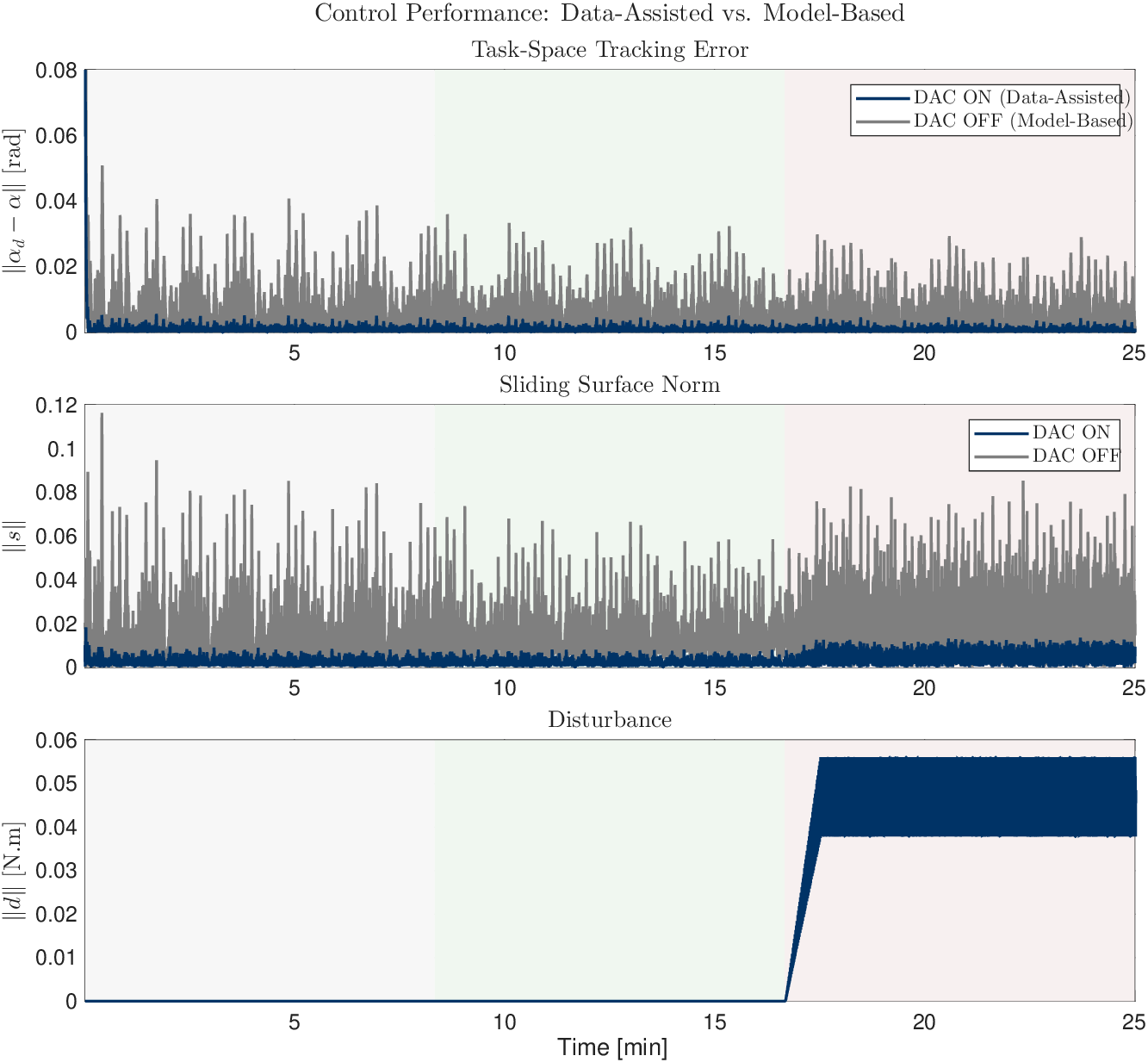}
         \caption{Control error, sliding surface error, and disturbance}
         \label{fig:DAC_Performance}
     \end{subfigure}
     \hfill
     \begin{subfigure}[b]{0.48\textwidth}
         \centering
         \includegraphics[width=\textwidth]{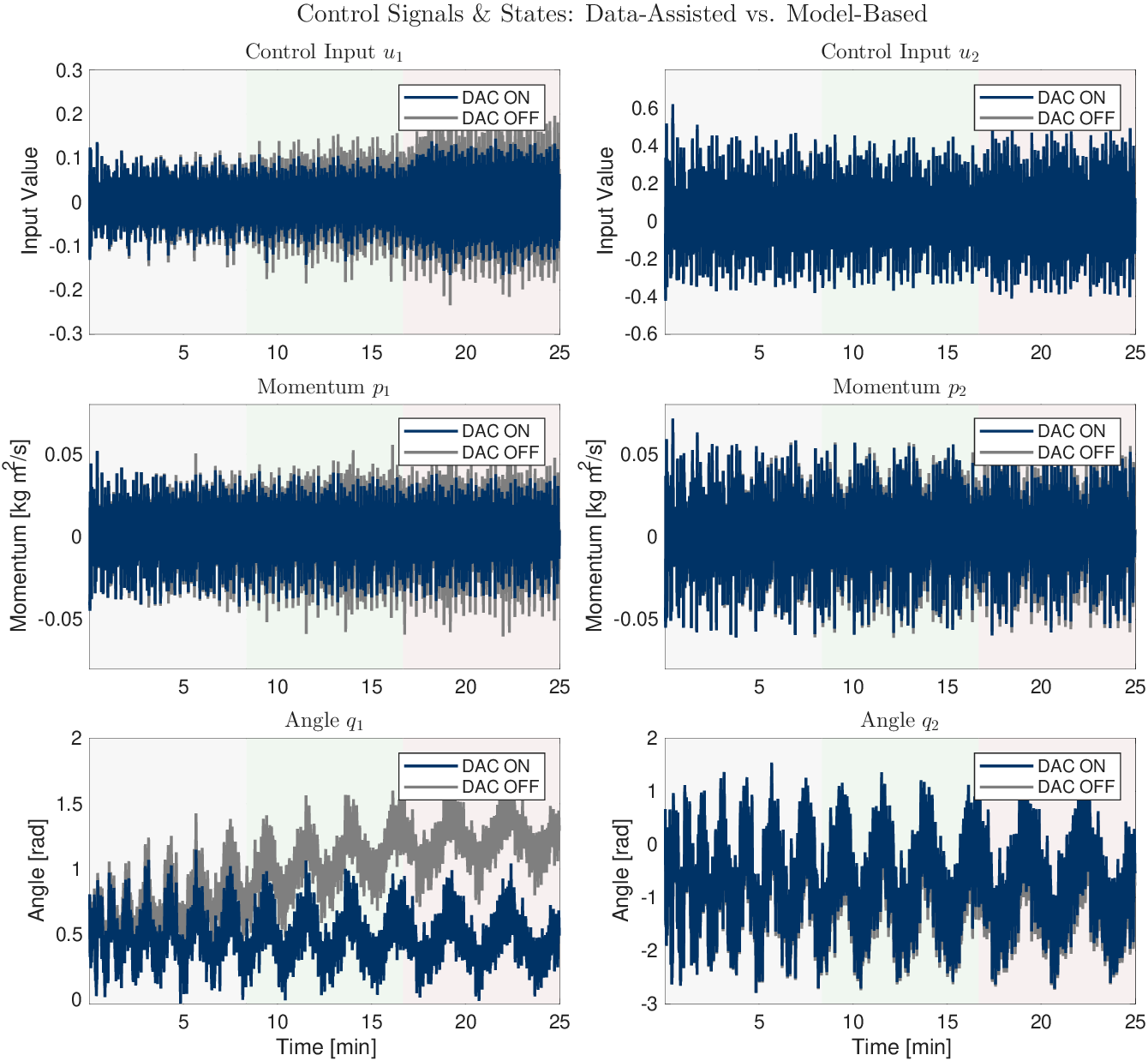}
         \caption{Control input and states}
         \label{fig:DAC_Performance_elements}
     \end{subfigure}
        \caption{DAC Performance comparison with Model-Based Control, in case of uncertainty and presence of disturbance}
        \label{fig:DAC}
\end{figure}

The simulation results, presented in Figures~\ref{fig:DNN} through \ref{fig:DAC}, validate the performance of the proposed DAC architecture. The evaluation is structured in three distinct phases, each lasting eight minutes and marked by colored regions in the figures. The scenario begins with the introduction of state-dependent linear uncertainty, progresses to include additional state-dependent nonlinear uncertainties, and concludes with the application of a significant external disturbance, $d(t)$, to the RHS. All figures demonstrate the change in estimated matrices by the introduced disturbance in the final phase, as noted previously.

Figure~\ref{fig:DNN_Performance} illustrates the performance of the DNN-based identification and the RHS controller in rejecting uncertainty and disturbances. The successful tracking of the individual elements of the true underlying matrices, $\tilde{B}$ and $\tilde{D}$, by the learned estimates is detailed in Figure~\ref{fig:DNN_Performance_elements}. To highlight the generalizability of the DNN, Figure~\ref{fig:DNN_Gen} provides a magnified view of this tracking performance shortly after the introduction of complex nonlinearities, showcasing the network's capability to adapt and learn the new dynamics. Quantitatively, the relative L2 norm of the estimation error for the $\tilde{B}$ and $\tilde{D}$ matrices was reduced to 63\% and 30\%, respectively, by the end of the second phase.

The adaptive nature of the control system is evident in Figure~\ref{fig:DAC_gains}, which displays the RHS integrator gains and the critical stability gain, $c$. These gains are updated online according to the stability condition in (48), with a selected safety factor of $T_d=2$. This choice establishes a significant robustness margin, ensuring that the potentially aggressive learning behavior of the DNN does not compromise the stability of the overall system. A key outcome, shown in the bottom plot of Figure~\ref{fig:DAC_gains}, is the considerable reduction in the residual uncertainty bound, $\chi$, which is passed to the LHS controller.

Figure~\ref{fig:DAC} provides a direct comparison of the proposed DAC controller against a purely model-based controller. The results demonstrate a significant improvement in both performance and robustness. The DAC framework achieves a marked reduction in the end-effector attitude tracking error and the sliding surface norm, and it completely rejects the high-impact disturbance introduced in the final phase. The DAC controller achieved a 77.5\% relative reduction in tracking error while requiring only a 16\% increase in control effort. This indicates that for the same level of performance, the DAC system consumes considerably less energy, validating its superior efficiency.

\section{Conclusion}\label{sec:conclusion}
This paper introduced a generic Data-Assisted Control architecture within the port-Hamiltonian framework, establishing a principled and physically interpretable method for controlling complex robotic systems. By introducing a new physical observable, the virtual port, it successfully decouples the system's well-modeled conservative dynamics from the uncertain, nonlinear effects of actuation, dissipation, and external disturbances. A robust, model-based sliding mode controller was designed for the conservative LHS to guarantee stability, while a decentralized integrator controller was employed as the inner loop, and a Deep Neural Network was employed to learn the algebraic mapping of the RHS.

A formal Lyapunov analysis confirmed the stability of the entire closed-loop system, with a strict time-scale separation ensuring that the learning component cannot destabilize the primary dynamics. Simulations of a free-floating space manipulator validated the architecture's effectiveness. The results demonstrated significant improvements in tracking performance and robustness compared to a purely model-based approach, achieving a 77.5\% reduction in tracking error while simultaneously handling complex uncertainties and rejecting large disturbances. This work confirms that by embedding physical priors, the DAC-pH framework offers a data-efficient, generalizable, and robust solution for the control of modern robotic systems operating in uncertain environments. Future work will explore the application of this framework to systems with more complex interaction dynamics and investigate the use of different learning algorithms for the RHS.

\section*{Declarations}
This work was not supported by any organization.

\bibliographystyle{unsrt}  
\bibliography{ref}

\begin{thebibliography}{10}

\bibitem{ellery2019tutorial}
Alex Ellery.
\newblock Tutorial review on space manipulators for space debris mitigation.
\newblock {\em Robotics}, 8(2):34, 2019.

\bibitem{papadopoulos2021robotic}
Evangelos Papadopoulos, Farhad Aghili, Ou~Ma, and Roberto Lampariello.
\newblock Robotic manipulation and capture in space: A survey.
\newblock {\em Frontiers in Robotics and AI}, 8:686723, 2021.

\bibitem{tsiotras2023spacecraft}
Panagiotis Tsiotras, Matthew King-Smith, and Lorenzo Ticozzi.
\newblock Spacecraft-mounted robotics.
\newblock {\em Annual Review of Control, Robotics, and Autonomous Systems}, 6(1):335--362, 2023.

\bibitem{alizadeh2024comprehensive}
Mohammad Alizadeh and Zheng~H Zhu.
\newblock A comprehensive survey of space robotic manipulators for on-orbit servicing.
\newblock {\em Frontiers in Robotics and AI}, 11:1470950, 2024.

\bibitem{van2006port}
Arjan Van Der~Schaft.
\newblock {H}amiltonian systems: an introductory survey.
\newblock In {\em Proceedings of the international congress of mathematicians}, volume~3, pages 1339--1365. Marta Sanz-Sole, Javier Soria, Juan Luis Verona, Joan Verdura, Madrid, Spain, 2006.

\bibitem{duindam2009modeling}
Vincent Duindam, Alessandro Macchelli, Stefano Stramigioli, and Herman Bruyninckx.
\newblock {\em Modeling and control of complex physical systems: the {H}amiltonian approach}.
\newblock Springer Science \& Business Media, 2009.

\bibitem{duong2024port}
Thai Duong, Abdullah Altawaitan, Jason Stanley, and Nikolay Atanasov.
\newblock Port-{H}amiltonian neural {ODE} networks on {L}ie groups for robot dynamics learning and control.
\newblock {\em IEEE Transactions on Robotics}, 40:3695--3715, 2024.

\bibitem{sprangers2015}
Olivier Sprangers, Robert Babuška, Subramanya~P. Nageshrao, and Gabriel A.~D. Lopes.
\newblock Reinforcement learning for {H}amiltonian systems.
\newblock {\em IEEE Transactions on Cybernetics}, 45(5):1017--1027, 2015.

\bibitem{nageshrao2014}
S.P. Nageshrao, G.A.D. Lopes, D.~Jeltsema, and R.~Babuška.
\newblock Passivity-based reinforcement learning control of a 2-dof manipulator arm.
\newblock {\em Mechatronics}, 24(8):1001--1007, 2014.

\bibitem{secchi2007control}
Cristian Secchi, Cesare Fantuzzi, and Stefano Stramigioli.
\newblock {\em Control of interactive robotic interfaces: A {H}amiltonian approach}.
\newblock Springer, 2007.

\bibitem{angerer2017port}
Martin Angerer, Selma Musi{\'c}, and Sandra Hirche.
\newblock {H}amiltonian based control for human-robot team interaction.
\newblock In {\em 2017 IEEE International Conference on Robotics and Automation (ICRA)}, pages 2292--2299. IEEE, 2017.

\bibitem{eslami2024data}
Mostafa Eslami and Afshin Banazadeh.
\newblock Data-assisted control: A framework development by exploiting nasa generic transport platform.
\newblock {\em International Journal of Robust and Nonlinear Control}, 34(3):1898--1920, 2024.

\bibitem{eslami2023sequential}
Mostafa Eslami and Afshin Banazadeh.
\newblock Sequential data-assisted control in flight.
\newblock {\em arXiv}, 2023.

\bibitem{eslami2025generalization}
Mostafa Eslami and Maryam Babazadeh.
\newblock On the generalization of data-assisted control in {H}amiltonian systems (dac-ph).
\newblock {\em arXiv preprint arXiv:2506.07079}, 2025.

\bibitem{papadopoulos1993dynamic}
Evangelos Papadopoulos and Steven Dubowsky.
\newblock {\em Dynamic Singularities in Free-floating Space Manipulators}, pages 77--100.
\newblock Springer US, Boston, MA, 1993.

\bibitem{xu1991dynamic}
Yangsheng Xu and Heung-Yeung Shum.
\newblock Dynamic control of a space robot system with no thrust jets controlled base.
\newblock Technical report, 1991.

\bibitem{mittal1994nonlinear}
Manoj Mittal, C-H Chuang, and Jer-Nan Juang.
\newblock Nonlinear control of space manipulators with model uncertainty.
\newblock In {\em Guidance, Navigation, and Control Conference}, page 3654, 1994.

\bibitem{papadopoulos1993nonholonomic}
Evangelos~G Papadopoulos.
\newblock Nonholonomic behavior in free-floating space manipulators and its utilization.
\newblock In {\em Nonholonomic Motion Planning}, pages 423--445. Springer, 1993.

\bibitem{spong2006robot}
Mark~W Spong, Seth Hutchinson, Mathukumalli Vidyasagar, et~al.
\newblock {\em Robot modeling and control}, volume~3.
\newblock Wiley New York, 2006.

\bibitem{nanos2012cartesian}
Kostas Nanos and Evangelos Papadopoulos.
\newblock On cartesian motions with singularities avoidance for free-floating space robots.
\newblock In {\em 2012 IEEE International Conference on Robotics and Automation}, pages 5398--5403. IEEE, 2012.

\bibitem{reinhart2017hybrid}
René~Felix Reinhart, Zeeshan Shareef, and Jochen~Jakob Steil.
\newblock Hybrid analytical and data-driven modeling for feed-forward robot control.
\newblock {\em Sensors}, 17(2), 2017.

\bibitem{umetani1989resolved}
Y.~Umetani and K.~Yoshida.
\newblock Resolved motion rate control of space manipulators with generalized jacobian matrix.
\newblock {\em IEEE Transactions on Robotics and Automation}, 5(3):303--314, 1989.

\bibitem{wang2016adaptive}
Hanlei Wang.
\newblock Adaptive control of robot manipulators with uncertain kinematics and dynamics.
\newblock {\em IEEE Transactions on Automatic Control}, 62(2):948--954, 2016.

\bibitem{eslami2016integrity}
Mostafa Eslami and Amin Nobakhti.
\newblock Integrity of lti time-delay systems.
\newblock {\em IEEE Transactions on Automatic Control}, 61(2):562--567, 2016.

\bibitem{karniadakis2021physics}
George~Em Karniadakis, Ioannis~G Kevrekidis, Lu~Lu, Paris Perdikaris, Sifan Wang, and Liu Yang.
\newblock Physics-informed machine learning.
\newblock {\em Nature Reviews Physics}, 3(6):422--440, 2021.

\bibitem{raissi2019physics}
Maziar Raissi, Paris Perdikaris, and George~E Karniadakis.
\newblock Physics-informed neural networks: A deep learning framework for solving forward and inverse problems involving nonlinear partial differential equations.
\newblock {\em Journal of Computational physics}, 378:686--707, 2019.

\bibitem{beucler2021enforcing}
Tom Beucler, Michael Pritchard, Stephan Rasp, Jordan Ott, Pierre Baldi, and Pierre Gentine.
\newblock Enforcing analytic constraints in neural networks emulating physical systems.
\newblock {\em Physical review letters}, 126(9):098302, 2021.

\bibitem{lutter2019deep}
Michael Lutter, Christian Ritter, and Jan Peters.
\newblock Deep lagrangian networks: Using physics as model prior for deep learning.
\newblock {\em arXiv preprint arXiv:1907.04490}, 2019.

\bibitem{greydanus2019hamiltonian}
Samuel Greydanus, Misko Dzamba, and Jason Yosinski.
\newblock Hamiltonian neural networks.
\newblock {\em Advances in neural information processing systems}, 32, 2019.

\bibitem{chen2020physics}
Yuyao Chen, Lu~Lu, George~Em Karniadakis, and Luca Dal~Negro.
\newblock Physics-informed neural networks for inverse problems in nano-optics and metamaterials.
\newblock {\em Optics express}, 28(8):11618--11633, 2020.

\bibitem{blaise2023}
James Blaise and Michael C.~F. Bazzocchi.
\newblock Space manipulator collision avoidance using a deep reinforcement learning control.
\newblock {\em Aerospace}, 10(9), 2023.

\bibitem{slotine1987on}
Jean-Jacques~E. Slotine and Weiping Li.
\newblock On the adaptive control of robot manipulators.
\newblock {\em The International Journal of Robotics Research}, 6(3):49--59, 1987.

\bibitem{khatib1985real}
O.~Khatib.
\newblock Real-time obstacle avoidance for manipulators and mobile robots.
\newblock In {\em Proceedings. 1985 IEEE International Conference on Robotics and Automation}, volume~2, pages 500--505, 1985.

\bibitem{ljung1998system}
Lennart Ljung.
\newblock System identification.
\newblock In {\em Signal analysis and prediction}, pages 163--173. Springer, 1998.

\bibitem{eslami2019robust}
Mostafa Eslami, Cheng~Siong Chin, and Amin Nobakhti.
\newblock Robust modeling, sliding-mode controller, and simulation of an underactuated rov under parametric uncertainties and disturbances.
\newblock {\em Journal of Marine Science and Application}, 18(2):213--227, 2019.

\end{thebibliography}

\end{document}